\documentclass[10pt, twocolumn, aps, prc, longbibliography, nofootinbib, floatfix, superscriptaddress]{revtex4-1}
\usepackage{graphicx, bm, xcolor}
\usepackage{hyperref}
\hypersetup{colorlinks, linkcolor={blue},citecolor={blue},urlcolor={blue}}
\usepackage{amsmath}
\usepackage{amssymb}
\usepackage[normalem]{ulem}
\usepackage{braket}
\usepackage{subfigure}
\usepackage{float}

\newcommand{\vect}[1]{\bm{\mathrm{#1}}} 
\newcommand{\kv}{\vect{k}}
\newcommand{\qv}{\vect{q}}
\newcommand{\hatvect}[1]{{\vect{\hat{#1}}}}
\newcommand{\vech}[1]{\vec{#1}} 
\newcommand{\hatvech}[1]{\hat{#1}}
\newcommand{\sigmabar}{\overline{\sigma}}
\newcommand{\Eq}[1]{Eq.~(\ref{#1})}
\newcommand{\calV}{\mathcal{V}}

\newcommand{\maxi}{\text{max}}
\newcommand{\fm}{\text{fm}}
\newcommand{\fmi}{\text{fm}{}^{-1}}

\newcommand{\Vlowk}{\ensuremath{V_{\text{low-}k}}}
\newcommand{\kF}{k_\text{F}} 

\newcommand{\new}[1]{{\color{red}#1}}

\usepackage{tikz}
\usetikzlibrary{decorations.pathmorphing}
\tikzset{snake it/.style={decorate, decoration=snake}}
\usetikzlibrary{positioning}
\tikzstyle{block} = [draw, fill=gray, rectangle, minimum height=4.6em, minimum width=0.5em]
\begin{document}

\title{Induced three-neutron interactions with low cutoffs for dilute neutron matter }
\author{Viswanathan Palaniappan}
\email{viswanathan@physics.iitm.ac.in}
\affiliation{Department of Physics, Indian Institute of Technology
  Madras, Chennai - 600036, India}
\affiliation{Universit\'e Paris-Saclay, CNRS-IN2P3, IJCLab, 91405
  Orsay, France}
\author{S. Ramanan}
\email{suna@physics.iitm.ac.in}
\affiliation{Department of Physics, Indian Institute of Technology
  Madras, Chennai - 600036, India}
\author{Michael Urban}
\email{michael.urban@ijclab.in2p3.fr}
\affiliation{Universit\'e Paris-Saclay, CNRS-IN2P3, IJCLab, 91405
	Orsay, France}
\affiliation{Department of Physics, Indian Institute of Technology
  Madras, Chennai - 600036, India}
\begin{abstract} 
The properties of dilute neutron matter are mostly determined by the $s$-wave two-body (2N) interaction, 
while three-body (3N) interactions are suppressed by the Pauli principle. 
In a previous work, we showed that it can be advantageous to use renormalization group based effective interactions with cutoffs scaled with the Fermi momentum, especially at low densities. 
In that case, induced 3N interactions may become important.
In this work, we compute the 3N interaction induced by the similarity renormalization group flow of 
the $s$-wave 2N interaction.
We work in the momentum-space hyperspherical partial wave basis and investigate its convergence
properties. Then we study the effect of the induced 3N interaction on the equation of state of dilute neutron matter.
 We observe that the cutoff dependence of the equation of state is strongly reduced when the effect of induced 3N interaction is included.

\end{abstract}
\maketitle
\section{Introduction}

The superfluid unbound neutrons in the inner crust of neutron stars are crucial
for understanding their thermal evolution and the phenomenon of the observed pulsar glitches~\cite{Chamel2008}.
In addition, the neutron gas is strongly correlated, although dilute.
Our objective is to obtain reliable dilute neutron matter properties, 
such as the equation of state (EoS) and the pairing gap from microscopic calculations.
This would allow us to better constrain the energy density functionals, which are used in calculations of the inner crust, but do not have the correct behavior at relevant low densities \cite{Yang2016,Gupta2024,Grams2024}, and pairing-related quantities such as the superfluid density, the heat capacity, and neutrino emission rates \cite{Sedrakian2019}.

The properties of a very dilute Fermi gas are determined by the $s$-wave scattering length $a$ \cite{Lee1957, FetterWalecka}. 
However, for neutron matter, 
the finite range of the interaction is also very important 
and as a result, one usually uses phenomenological two-body (2N) interactions (such as AV18 \cite{Wiringa1995})
as input.
Such interactions have a strong repulsive core and 
hence are not amenable to perturbative calculations.
In fact, it is well-known that the nuclei are not even bound in the Hartree-Fock (HF) theory when these 
phenomenological interactions are used.
Instead, one has to do resummations, as, e.g., in the Brueckner theory \cite{HjorthJensen1995, Baldo1999}.
Even with the modern chiral interactions \cite{Entem2003,Entem2015},
which are soft, 
one still has to resum ladder diagrams in order to obtain
the correct EoS of neutron matter, especially in the very low-density regime \cite{Hebeler2013}.

An alternative is to use the so-called softened interactions such as 
the renormalization group based low-momentum interactions (\Vlowk) 
or the similarity renormalization group (SRG) interactions \cite{Bogner2003,Bogner2007,Bogner2010}.
In addition to decoupling the low- and high-momentum regions and softening the interactions,
these have an important additional property: 
as one decreases the renormalization scale (cutoff) $\Lambda$ towards zero,
the 2N matrix element $V_{^1S_0}(0,0)$ flows to the scattering length $a$.
Taking the cutoff $\Lambda \sim 2 \kF$ makes the effective
interaction perturbative as discussed in \cite{Bogner2005}.

It has been demonstrated in~\cite{Ramanan2018} that 
the correct low-density limit for the superfluid critical
temperature is obtained with a density-dependent cutoff $\Lambda=f \kF$, 
where $\kF$ is the Fermi momentum and $f$ is a scale factor, 
due to the effective resummation of the ladders, inherent in the
flow towards the scattering length as $\Lambda \to 0$ \cite{Bogner2005}.
In our previous work \cite{Palaniappan2023}, we have shown that 
it is possible to obtain the correct EoS of neutron matter at very low densities  
using a 2N \Vlowk{} interaction with a density-dependent cutoff
and a scale factor $f \sim 1.5-2.5$.
In addition, the Bogoliubov many-body perturbation theory (BMBPT) corrections 
to the Hartree-Fock-Bogoliubov (HFB) ground state energy obtained using a density-dependent cutoff 
have better convergence compared to those obtained using a fixed cutoff. 
However, the HFB+BMBPT results in \cite{Palaniappan2023} show a small residual 
cutoff dependence, which indicates missing corrections. 
It is known that as one decreases the cutoff, three- and higher-body
interactions are induced \cite{Bogner2010}. 
However, these were not included in our previous work, since it is not straight-forward to extend $\Vlowk$ to higher-body interactions.

The SRG presents an efficient method to consistently evolve two- and higher-body interactions. Because of the Pauli principle, the 3N contact term, which should be dominant in symmetric matter at very low density, is absent in pure neutron matter \cite{Hebeler2010}. The leading ``bare'' (or ``genuine'') 3N force for three neutrons is the two-pion exchange contribution which starts to contribute only at higher densities. Therefore, in the present paper, we neglect the bare 3N force and consider only the one induced by the SRG flow. Furthermore, we restrict ourselves to the 3N interactions induced by the 2N $s$-wave interaction, since this is by far the dominant channel at low densities.

Concerning the practical implementation, one possible choice for the 3N momentum space basis is the Jacobi partial-wave basis which has been 
used in the SRG-evolved 3N interactions in the study of both finite nuclei as well as infinite matter \cite{Hebeler2012, Hebeler2013, Hebeler2021}.
An alternative momentum space basis for the 3N interaction is the hyperspherical harmonic 
or hyperspherical partial-wave (HPW) basis, as was first demonstrated and applied to the triton in \cite{Wendt2013}. 

In Section \ref{sec:formalism}, we obtain the induced 3N interaction in the momentum space HPW basis using the SRG formalism.
In Sec. \ref{sec:results}, we discuss our results for the induced 3N matrix elements, their convergence in the HPW basis, and their effect on the EoS of dilute neutron matter, in particular the cutoff dependence.
In Sec. \ref{sec:summary}, we summarize our current work and present an outlook for possible future studies.
Technical details are provided in the appendices.

\section{Formalism}\label{sec:formalism}

\subsection{Similarity Renormalization Group}
The SRG flow equation is given by \cite{Bogner2007}
\begin{align}
    \frac{dH_s}{ds} &= [\eta_s,H_s]\,,
\end{align}
where $\eta_s$ is the generator of the unitary transformation, $H_s$ is the Hamiltonian, and $s$ is the flow parameter. The flow parameter $s$ is usually replaced with the decoupling scale $\lambda\equiv s^{-1/4}$. It is well known that the flow equation induces higher-body interactions as one evolves in $s$ \cite{Bogner2007,Bogner2010}.

\begin{figure}
\centering
\includegraphics{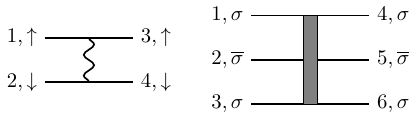}
\caption{Schematic representation of the two- and three-body terms in the Hamiltonian ($\sigmabar = -\sigma$).}\label{fig:Hamiltonian}
\end{figure}

Since we are interested in dilute neutron matter, we concentrate on the 3N interaction induced by the 2N $s$-wave ($^1S_0$) interaction 
and neglect 3N interactions induced by higher partial waves.
Further, the bare 3N interaction as well as induced 4N (and higher) interactions are also neglected.
In an induced 3N interaction, one of the three particles interacts consecutively with the other two particles via 2N interactions. If the 2N interactions are restricted to the $s$-wave channel, then this particle must have an opposite spin with respect to the other two particles. And since the $s$-wave interaction does not flip the spins, the incoming and outgoing particles for the induced 3N interaction must have the same spins. Fig.~\ref{fig:Hamiltonian} shows a schematic illustration of the 2N and 3N terms in the Hamiltonian.

Therefore, we consider the Hamiltonian to be of the following form
\begin{align}
\label{eq:hamiltonian}
{H}_s 
&= {T} + {V}_s + {W}_s
\nonumber\\
&= 
\sum_{\vect{p}\sigma} \varepsilon^{}_{p} a^{\dagger}_{\vect{p}\sigma} a^{}_{\vect{p}\sigma}
+ \sum_{\substack{1\dots4}} 
V_{1234}(s) 
a^{\dagger}_{1\uparrow} a^{\dagger}_{2\downarrow}
a^{}_{4\downarrow} a^{}_{3\uparrow}
\nonumber\\
&\quad
+ \frac{1}{8}
\sum_{\substack{1\dots6\\\sigma}}
W_{123456}(s)
a^{\dagger}_{1\sigma} a^{\dagger}_{2\sigmabar} a^{\dagger}_{3{\sigma}}
a^{}_{6{\sigma}} a^{}_{5\sigmabar} a^{}_{4\sigma}\,.
\end{align}
Here, $\varepsilon_p = p^2/2$ is the kinetic energy of a free neutron with momentum $\vect{p}$, in units with $\hbar=c=m=1$. $a^{}_{\vect{p}\sigma}$ and 
$a^{\dagger}_{\vect{p}\sigma}$ are, respectively, the annihilation and creation operators of a neutron with momentum $\vect{p}$ and spin $\sigma$. The 
numbers $i=1,2,3,\dots$ are short-hand notations for the momenta $\vect{p}_i$ and $\sigmabar = -\sigma$. 
$V_{1234}$ are the matrix elements of the 2N $s$-wave interaction, 
and $W_{123456}$ are the matrix elements of the induced 3N interaction, 
antisymmetrized with respect to the exchange of in- and outgoing particles having the same spin, 
i.e., with respect to $1\leftrightarrow 3$ and $4\leftrightarrow 6$ 
(this is the reason for the factor $1/8$ in Eq.~\eqref{eq:hamiltonian}). 
Since we do not include any initial 3N interactions, we set $W_{123456}(0)=0$.

Using the common choice for the generator $\eta_s=[T, H_s]$, we obtain the flow equation for the two- and three-body matrix elements,
\begin{widetext}
\begin{align}
    \frac{d}{ds} V_{1234}
    =& - (\varepsilon_{12}-\varepsilon_{34})^2
    V_{1234}
    + \sum_{56} (\varepsilon_{12}+\varepsilon_{34}-2\varepsilon_{56})
    V_{1256} V_{5634}\,,\label{eq:srg2sp}
\\
    \frac{d}{ds} W_{123456} 
    =& - (\varepsilon_{123}-\varepsilon_{456})^2 W_{123456}
\nonumber\\
& 
      + \sum_{789} (\varepsilon_{123}+\varepsilon_{456}-2\varepsilon_{789})
    \Big[
        2(A^{(13)}V^{(12)})_{123789} (V^{(32)}A^{(13)})_{789456}
        \nonumber\\
        &\qquad+(A^{(13)}V^{(12)})_{123789} W_{789456}
        +W_{123789}(V^{(12)}A^{(13)})_{789456}
        +\frac{1}{4}W_{123789}W_{789456}
    \Big]\,.\label{eq:srg3sp}
\end{align}
\end{widetext}
where $\varepsilon_{ij}=\varepsilon_{i}+\varepsilon_{j}$, and $\varepsilon_{ijk}=\varepsilon_{i}+\varepsilon_{j}+\varepsilon_{k}$.
In \Eq{eq:srg3sp}, we used the following definitions and notations: 
$A^{(ij)}=I-P^{(ij)}$ is proportional to the antisymmetrizer with respect to particles $i$ and $j$. To be specific, $I_{1231'2'3'}=\delta_{11'}\delta_{22'}\delta_{33'}$ is the 3N identity operator and $P^{(13)}_{1231'2'3'}=\delta_{13'}\delta_{22'}\delta_{31'}$ is the permutation operator of particles 1 and 3. Notations of the form $(AB)_{1231'2'3'}$ should be understood as operator products $\sum_{1''2''3''}A_{1231''2''3''}B_{1''2''3''1'2'3'}$. $V^{(ij)}$ denotes the 2N interaction between particles $i$ and $j$ ``embedded'' in the 3N space. For example,
\begin{equation}\label{eq:V12I3_1}
    V^{(12)}_{1231'2'3'} = V_{121'2'}\delta_{33'}\,.
\end{equation}
Notice that $V^{(12)}$ includes a Kronecker delta for the spectator particle 3. 

When discussing the different terms on the right-hand side of \Eq{eq:srg3sp}, we will call the first term the linear term, the second term (second line of \Eq{eq:srg3sp}) the $VV$ term, and the last three terms (third line of \Eq{eq:srg3sp}) the $VW$, $WV$, and $WW$ terms, respectively.

\subsection{Two-body flow equation}

For simplicity, we have written the formal equations (\ref{eq:hamiltonian}) - (\ref{eq:srg3sp}) for a finite volume $\calV$ in which the momenta are discrete, but now we can go to the continuum limit.
Let us first look at the 2N flow equation. 
(The normalization conventions for the 2N states are given in Appendix \ref{appendix:2N_basis}.)
The 2N matrix elements are given by 
\begin{equation}\label{eq:V2Ncm}
    V_{121'2'} = \frac{\delta_{\vect{P}\vect{P}'}}{\calV} 4 \pi V(k,k').
\end{equation}
where $\mathbf{P}=\mathbf{p}_1+\mathbf{p}_2$, $\mathbf{P}'=\mathbf{p}_1'+\mathbf{p}_2'$ are the total momenta, $\mathbf{k}=(\mathbf{p}_1-\mathbf{p}_2)/2$, $\mathbf{k}'=(\mathbf{p}_1'-\mathbf{p}_2')/2$ are the momenta in the center-of-momentum frame, and $V(k,k')$ is the $s$-wave matrix element. 
The 2N flow equation \Eq{eq:srg2sp} becomes
\begin{align}\label{eq:SRG2}
    \frac{d}{ds}{V}(k,k') 
    =& - (k^2-k'^2)^2 {V}(k,k') 
    \nonumber \\
    &+ \frac{2}{\pi} \int d k'' k''^2 (k^2+k'^2-2 k''^2) 
    \nonumber \\
    &\times {V}(k,k''){V}(k'',k').
\end{align}
The coefficients in the linear term make the flow equation a stiff differential equation. 
In order to reduce the stiffness, 
the exponential time differencing (ETD) method \cite{Cox2002} is used
where the linear part is solved exactly 
by multiplying both sides with an exponential integrating factor,  
followed by the Runge-Kutta method to solve the differential equation.

\subsection{Three-body flow equation}
Before looking at the 3N flow equation, we will introduce some definitions for working in the momentum representation in the 3N space.
To separate center-of-mass and relative motion in the three-body case, 
it is useful to introduce the total momentum $\vect{P}$ and the Jacobi momenta $\vect{k}$ and $\vect{q}$. 
There are actually six possible definitions of the Jacobi momenta, obtained by permuting the three particle labels. 
As an example, we show in Fig.~\ref{fig:Jacobi_momenta} the definition of the third Jacobi system $\mathcal{J}^{(3)}$. 
See Appendix \ref{appendix:jacobi_basis} for more details.

\begin{figure}
\centering
\includegraphics{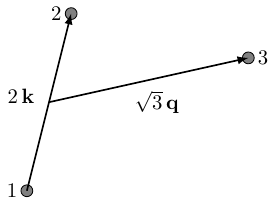}
\caption{Jacobi momenta in the Jacobi system $\mathcal{J}^{(3)}$} \label{fig:Jacobi_momenta}
\end{figure}

The 3N matrix elements in the Jacobi basis are given by
\begin{equation}
    W_{1231'2'3'}
    = \frac{\delta_{\vect{P}\vect{P'}}}{J\mathcal{V}^2} \braket{\vect{k}\vect{q}|W|\vect{k}'\vect{q}'}.
\end{equation}
The factor $\mathcal{V}^{-2}$ comes from taking the continuum limit for the relative momenta, and $J$ is the Jacobian of the transformation (cf. Appendix \ref{appendix:jacobi_basis}).

In the third Jacobi basis $\mathcal{J}^{(3)}$, the embedded matrix elements of $V^{(12)}$ (cf. Eq.~\eqref{eq:V12I3_1}) are given by
\begin{align}\label{eq:v2n_jacobi}
    \braket{\vect{k}\vect{q}|V^{(12)}|\vect{k}'\vect{q}'} = 4\pi V(k,k')\; (2\pi)^3\delta(\vect{q}-\vect{q}'),
\end{align}
where the delta function corresponds to the spectator particle.

As a generalization of the 2N partial-wave expansion, one can define the HPW basis for the 3N space. In addition to the angular momentum quantum numbers $l_1m_1$ and $l_2m_2$ describing the dependence on the angles of $\vect{k}$ and $\vect{q}$, there is a grand orbital momentum $L$ that characterizes the dependence on the relative magnitude of $k$ and $q$, and the hypermomentum $K=\sqrt{k^2+q^2}$. After coupling $l_1m_1$ and $l_2m_2$ to the total angular momentum $lm$, the set of discrete quantum numbers becomes $\{L,l,m,l_1,l_2\}\equiv [L]_c$. See Appendix \ref{appendix:HPW_basis} for details.

Transforming Eq.~\eqref{eq:v2n_jacobi} into the HPW basis, one obtains the embedded matrix elements that were used, e.g., in Ref.~\cite{Wendt2013} to solve the coupled two- and three-body SRG flow equations. They are given by 
\begin{align}
    \braket{K[L]_c|V^{(12)}|K'[L']_c} 
    = \delta^{l'}_{l}{}^{m'}_{m}{}^{l_1}_{0}{}^{l_2}_{l}{}^{l_1'}_{0}{}^{l_2'}_{l} 
    \, V_{lLL'}(K,K'),
\end{align}
where $\delta^{a'}_{a}{}^{b'}_b{}^{c'}_{c}{}^{\dots}_{\dots} \equiv \delta_{aa'}\delta_{bb'}\delta_{cc'}\dots$, and 
\begin{align}
    V_{lLL'}(K,K') 
    &= \frac{1}{2^{l+2}\sqrt{2}\pi} i^{L-L'} N_{L}^{l0}N_{L'}^{l0} \frac{K^l}{K'^{\,l+3}}
    \nonumber\\
    &\hspace{0.5cm}\times
    \int dx (1-x)^{l+\frac{1}{2}}(1+x)^{\frac{1}{2}}(1+x')^{\frac{1}{2}}
    \nonumber\\
    &\hspace{0.5cm}\times
    P_{\frac{L-l}{2}}^{(l+\frac{1}{2},\frac{1}{2})}(x)
    P_{\frac{L'-l}{2}}^{(l+\frac{1}{2},\frac{1}{2})}(x')
    \nonumber\\
    &\hspace{0.5cm}\times
    {V}\bigg(K\sqrt{\frac{1+x}{2}},K'\sqrt{\frac{1+x'}{2}}\bigg),
\end{align}    
where $x' = 1-\big(\frac{K}{K'}\big)^2 (1-x)$ \cite{Wendt-thesis}. In this equation, $N^{l_2l_1}_L$ is a normalization factor (see Appendix \ref{appendix:HPW_basis}) and $P^{(\alpha,\beta)}_n$ denotes a Jacobi polynomial. For $K\le K'$, the integral over $x$ runs from $-1$ to $1$ and can be computed using Gauss-Jacobi quadrature. For $K>K'$, one can use $V_{lLL'}(K,K') = V_{lL'L}(K',K)$.

Let us now look at the antisymmetrized 3N matrix elements in the HPW basis.
The antisymmetrizer $A^{(13)}$ conserves $K$, $L$, $l$, and $m$, but not $l_1$ and $l_2$. As explained in Appendix \ref{appendix:HPW_basis}, its matrix elements in the HPW basis can be written as $\mathcal{A}^{Ll}_{l_1l_2l'_1l'_2}(\theta_{13})$.
Inspecting the form of the inhomogeneous ($VV$) term of \Eq{eq:srg3sp}, one can see that the 3N matrix elements must be of the following form,
\begin{align}
    \braket{K[L]_c|W|K'[L']_c} &= \delta^{l'}_{l}{}^{m'}_{m}\mathcal{A}^{Ll}_{l_1l_20l}(\theta_{13})\mathcal{A}^{L'l}_{l_1'l_2'0l}(\theta_{13})
    \nonumber\\
    &\hspace{1cm}\times W_{lLL'}(K,K').
\end{align}

The flow equation for the 3N matrix elements then reduces to the following,
\begin{widetext}
\begin{align}\label{eq:SRG3_HPW}
    \frac{d}{ds}W_{lLL'}(K,K') = 
    &- (K^2-K'^2)^2 W_{lLL'}(K,K') 
    + [VV]_{lLL'}(K,K')
    + \sum_{L''} \int dK'' K''^5 (K^2+K'^2-2K''^2)
    \nonumber\\
    &\times \bigg[
    \mathcal{A}^{L''l}_{0l0l}(\theta_{13})\,\Big(V_{lLL''}(K,K'')W_{lL''L'}(K'',K')
    +W_{lLL''}(K,K'')V_{lL''L'}(K'',K')\Big)
    \nonumber\\
    &+\frac{1}{4} W_{lLL''}(K,K'')W_{lL''L'}(K'',K')\sum_{l_1''l_2''}(\mathcal{A}^{L''l}_{l_1''l_2''0l}(\theta_{13}))^2
    \bigg].
\end{align}
\end{widetext}
Unlike the last three ($VW$, $WV$, and $WW$) terms, we have not written the $VV$ term as a sum over $L''$. The reason for this is that in the $VV$ term, the $L''$ summation converges very slowly at low momenta. In fact, combining the spectator delta functions with the momentum-conserving delta functions of the 2N matrix elements $V$, one can see that in the $VV$ term of \Eq{eq:srg3sp}, the internal momenta $789$ can be expressed in terms of the external momenta $123456$, and there is actually no summation over intermediate states at all. Then, the $VV$ term can be computed from the following expression (refer to Appendix \ref{appendix:VV_term} for more details on the derivation):
\begin{widetext}
\begin{align}\label{eq:VV_HPW}
    [VV]_{lLL'}(K,K')
    =& - \frac{J}{2^l 16  \pi^2} i^{L-L'} N_L^{l0} N_{L'}^{l0} 
    \int dx (1-x)^{\frac{l+1}{2}} (1+x)^{\frac{1}{2}} P_{\frac{L-l}{2}}^{(l+\frac{1}{2},\frac{1}{2})}(x)
    \nonumber\\ &\times 
    \int dx' (1-x')^{\frac{l+1}{2}} (1+x')^{\frac{1}{2}} P_{\frac{L'-l}{2}}^{(l+\frac{1}{2},\frac{1}{2})}(x')
    \int dz P_l(z) \bigg(k^2+k'^2-\frac{5}{3}(q^2+q'^2)-\frac{8}{3}qq'z\bigg)
    \nonumber\\ &\times
    V\bigg(k,\sqrt{\frac{q^2+4q'^2+4qq'z}{3}}\bigg)
    V\bigg(\sqrt{\frac{4q^2+q'^2+4qq'z}{3}},k'\bigg),
\end{align}
\end{widetext}
where $k=K\cos{\alpha}$, $q=K\sin{\alpha}$, $\alpha = \arccos\sqrt{\vphantom{y^2}\smash{\frac{1+x}{2}}} = \allowbreak\tfrac{1}{2}\arccos x$, 
similarly for $k'$ and $q'$, and $P_l$ is a Legendre polynomial. 
The $x$ and $x'$ integrals can be computed using Gauss-Jacobi quadrature, 
while the $z$ integral can be computed using Gauss-Legendre quadrature.

Analogous to the 2N flow equation, the 3N flow equation
\eqref{eq:SRG3_HPW} can be solved using ETD techniques.

\section{Results and Discussion} \label{sec:results}
\subsection{2N and induced 3N matrix elements}
\begin{figure*}
    \centering
    {\includegraphics[width=\linewidth]{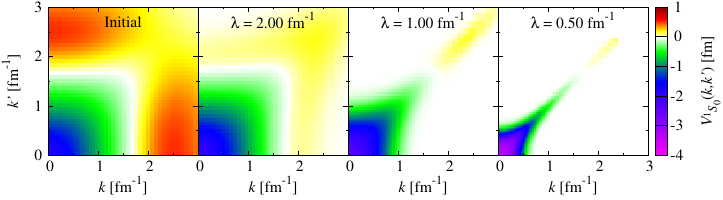}}
    {\includegraphics[width=\linewidth]{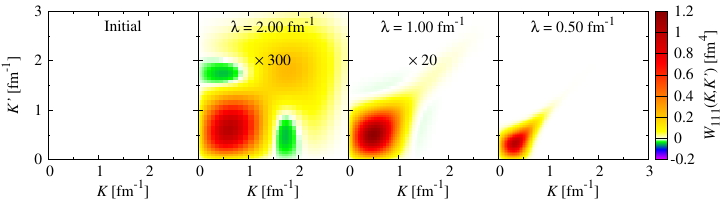}}
    \caption{Contour plots of the evolved 2N (upper panel) and induced 3N (lower panel) matrix elements for different SRG cutoffs. 
    For the 3N interaction, the channel $l=L=L'=1$ is chosen, which is the most important one, and the matrix elements in the second and third panels are multiplied by factors of 300 and 20, respectively, to make them visible.}
    \label{fig:SRG_flow}
\end{figure*}
The results presented in this section are obtained 
using the SRG-evolved N4LO chiral interaction \cite{Entem2015}.
Fig.~\ref{fig:SRG_flow} shows the matrix elements of $V_{^1S_0}$ 
and $W_{111}$ for different values of the SRG decoupling scale (``cutoff'') $\lambda=s^{-1/4}$. 
It can be seen that as the cutoff decreases, 
a 3N interaction is induced by the $^1S_0$ 2N interaction 
(as already mentioned, we neglect the 3N interaction induced by other channels of the 2N interaction, 
as well as the bare 3N interaction).
As expected, the two- and three-body matrix elements grow in magnitude at low momenta
and become band diagonal at large momenta. 
We see that the identification of $\lambda$ with a cutoff makes sense because the off-diagonal matrix elements are indeed very small for momenta that are larger than $\lambda$ (once $\lambda$ is smaller than the cutoff of the initial interaction).

\begin{figure}
    \centering
    {\includegraphics[width=\columnwidth]{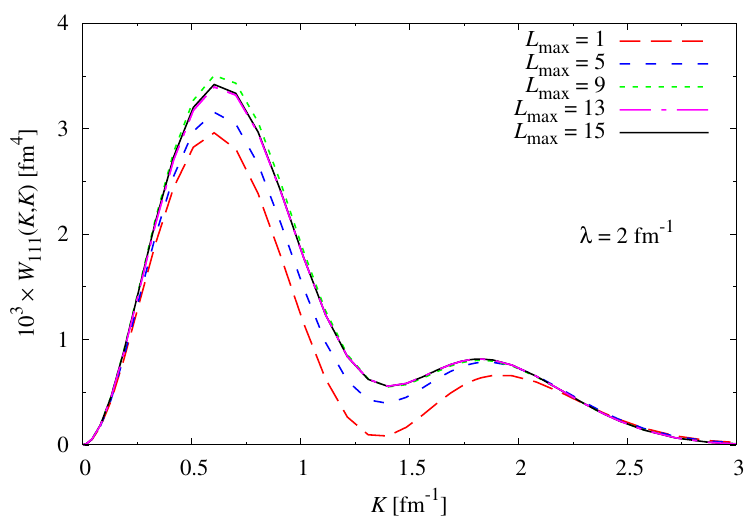}}
    \caption{Convergence of the diagonal matrix elements $W_{111}(K,K)$ (multiplied by a factor of $10^3$) with respect to the intermediate $L''$ summation in the 3N flow equation \Eq{eq:SRG3_HPW}.}
    \label{fig:mat_elem_conv}
\end{figure}

Since we truncate the intermediate $L''$ sum in \Eq{eq:SRG3_HPW} at some finite value $L_{\maxi}$,   
it is necessary to check that the matrix elements have converged.
Fig.~\ref{fig:mat_elem_conv} shows the convergence of the $W_{111}$ matrix elements. 
Convergence is reached when channels with $L''$ up to $L_{\maxi}=13$ are included in the summation.
We emphasize that this is only possible because we use \Eq{eq:VV_HPW} for the $VV$ term.
If instead of using \Eq{eq:VV_HPW}, 
the $VV$ term is computed using the embedded matrix elements along with the intermediate $L''$ summation, 
the convergence of the matrix elements at low momenta is much slower. 

\begin{figure}
\centering
\includegraphics[width=\columnwidth]{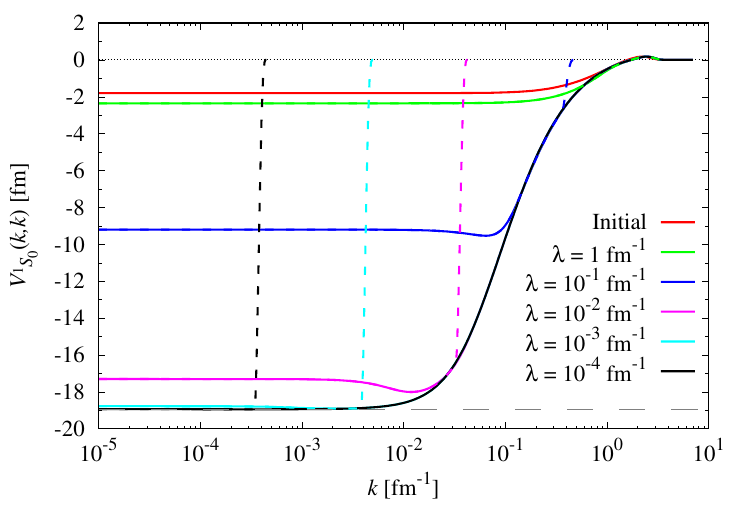}
\caption{Diagonal $^1S_0$ matrix elements for different SRG cutoffs. The dashed and solid lines correspond to matrix elements computed with and without readjusting the mesh, as explained in the text. The gray long dashed line corresponds to the $^1S_0$ scattering length $a=-18.95 \, \fm$ \cite{Entem2017}.} \label{fig:scattering_length}
\end{figure}

Since we use a density-dependent cutoff to obtain the correct low-density limit of the neutron matter EoS \cite{Palaniappan2023}, 
we have to flow the matrix elements to much lower cutoffs than is usually done.
Figure \ref{fig:scattering_length} shows the diagonal $^1S_0$ matrix elements for different cutoffs.
We note that they flow to $a$ (for $k\ll 1/a$) as one lowers the cutoff, similar to the $^1S_0$ $\Vlowk(0,0)$ matrix elements as shown in Fig. 15 in \cite{Bogner2010}.\footnote{This limit is in practice reached only at $\lambda\ll 1/a$ since $1/V_{^1S_0}(0,0) \approx 1/a-2\lambda/\pi$ \cite{Ramanan2018,Urban2021} as one can easily understand in the $\Vlowk$ case.}

While using a fine and large momentum mesh as in Fig.~\ref{fig:scattering_length} is feasible for demonstration purposes, 
it would be too time and memory consuming to do this for all partial waves, especially in the 3N case. 
Hence, we restrict the momentum mesh to the relevant range by readjusting it during the evolution. 
When the cutoff gets very small compared to the last point in the momentum mesh $k_\maxi$ (say, $\lambda = k_\maxi/5$), 
we change the momentum mesh in the following manner: 
(1) make the mesh denser at lower momenta ($k, k' \lesssim \lambda$), 
(2) cut the upper edge of the mesh ($k, k' > 3.8 \lambda$), 
where there is only a narrow tail of non-zero matrix elements along the diagonal, and 
(3) apply a smooth regulator at the upper edge of the new mesh to avoid numerical instabilities (due to step (2)) as we flow further down.
Since the tail being excluded is almost diagonal, step (2) has no effect on the results \cite{Bogner2007b}. This can be seen in Fig. \ref{fig:scattering_length} from the perfect agreement of the matrix elements for $k \lesssim 3.8 \lambda$ computed with (dashed lines) and without (solid lines) readjusting the mesh.

\subsection{Induced 3N contribution to the neutron-matter EoS}

\begin{figure}
    \centering
    {\includegraphics[width=\columnwidth]{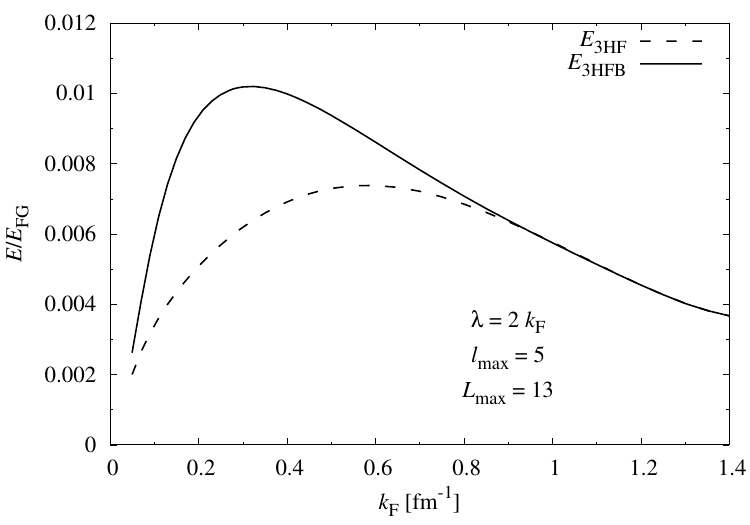}}
    \caption{Contribution of the induced 3N interaction to the energy of neutron matter (in units of the energy of the free Fermi gas) as a function of the Fermi momentum.}
    \label{fig:e3_hf_vs_hfb}
\end{figure}

As a first application of the induced 3N interaction, 
we will now discuss its contribution to the EoS of dilute neutron matter. 
In the simplest case, one can estimate its contribution in first-order perturbation theory by taking its expectation value in the HF ground state:
\begin{equation}\label{eq:HF3}
\varepsilon_{\text{3HF}}
=\frac{\braket{\text{HF}|W|\text{HF}}}{\mathcal{V}}
= \frac{1}{2\mathcal{V}} \sum_{p_1,p_2,p_3<\kF} W_{123123}\,.
\end{equation}
The integrals can be computed numerically using Monte\new{-}Carlo integration. 
The dashed line in Figure~\ref{fig:e3_hf_vs_hfb} shows the result 
in units of the ideal Fermi gas energy\footnote{The symbols $\varepsilon$ and $E$ denote, respectively, the energy density and the energy per particle, related by $E=\varepsilon/n$, where $n=\kF^3/(3\pi^2)$ is the density.} [$\varepsilon_{\text{FG}}=\kF^5/(10\pi^2)$]
as a function of the Fermi momentum, for the choice $\lambda = 2\kF$. 
As we can see, it is small and repulsive. 
For a bare 3N interaction, one would expect it to contribute only at high density, but for the induced interaction, it is completely different. 
The reason is the density-dependent cutoff: 
the smaller the $\kF$, the smaller $\lambda$ gets,
and the induced 3N matrix elements get bigger.
However, at very small $\kF$ values, the ratio $\varepsilon_{3\text{HF}}/\varepsilon_{\text{FG}}\propto \kF^4 \braket{W}$ starts to decrease because the matrix elements at momenta of order $\kF$ grow slower than $1/\lambda^4$.

However, in our previous work \cite{Palaniappan2023}, 
we have seen that, especially at very low density, the HF ground state is not the best starting point. 
There, the starting point was the HFB ground state to which perturbative corrections were added in BMBPT. 
Therefore, let us see how the contribution of the induced 3N interaction changes if we compute it in first-order perturbation theory by taking the expectation value in the HFB ground state instead of the HF one. 
In this exploratory study,
we do not include the effect of the induced 3N interaction on the HFB ground state 
or its BMBPT corrections (similar to Fig.~49 of Ref.~\cite{Hebeler2021}). 
Note that the 2N calculations presented in this and the next subsection are 
obtained with the full SRG evolved 2N interaction 
(including partial waves with $j\le6$ yields converged results 
for the density range $\kF \le 1.4\, \fmi$), 
while for the 3N calculations, we consider only the contribution of the $s$-wave-induced 3N interaction, as explained before.

In the notation of \cite{Palaniappan2023}, 
the expectation value of the 3N interaction in the HFB ground state is given by
\begin{align}\label{eq:HFB3}
\varepsilon_{\text{3HFB}}
&=\frac{\braket{\text{HFB}|W|\text{HFB}}}{\mathcal{V}}
\nonumber\\
&= \frac{1}{2\mathcal{V}} \sum_{\vect{p}_1 \vect{p}_2 \vect{p}_3}  \Big[ v^2_{1}\,v^2_{2}\,v^2_{3}\, W_{123123}
\nonumber\\
&\hspace{2cm}+ 2 \,u_{1}v_{1} \,u_{2}v_{2}\,v^2_{3} \, W_{1,-1,3,2,-2,3} \Big]\,.
\end{align}
In the limit of no pairing, the second term vanishes, 
the $v_i$ become step functions $\theta(\kF-p_i)$, 
and the first term reduces to the expression of the expectation value in the HF ground state, Eq.~\eqref{eq:HF3}. 
The result is shown in Fig.~\ref{fig:e3_hf_vs_hfb} as the solid line.

Comparing the HF and HFB expectation values,
we see that they are identical at $\kF\gtrsim 0.9\,\fmi$ 
when the gap becomes small compared to the Fermi energy 
(similar to the HF and HFB energies in the 2N case, cf. Fig.~4 in Ref.~\cite{Palaniappan2023}). 
While the first term in the right-hand side of \Eq{eq:HFB3} makes a similar contribution as in the HF case,
the second term gives a significant contribution at low densities ($\kF \lesssim 0.7\,\fmi$). 
The physical interpretation of this second term is the increase in energy 
due to the reduction of the pairing gap 
if one includes an additional contribution to the pairing interaction from the 3N interaction by closing one line.

\begin{figure}
    \centering
    {\includegraphics[width=\columnwidth]{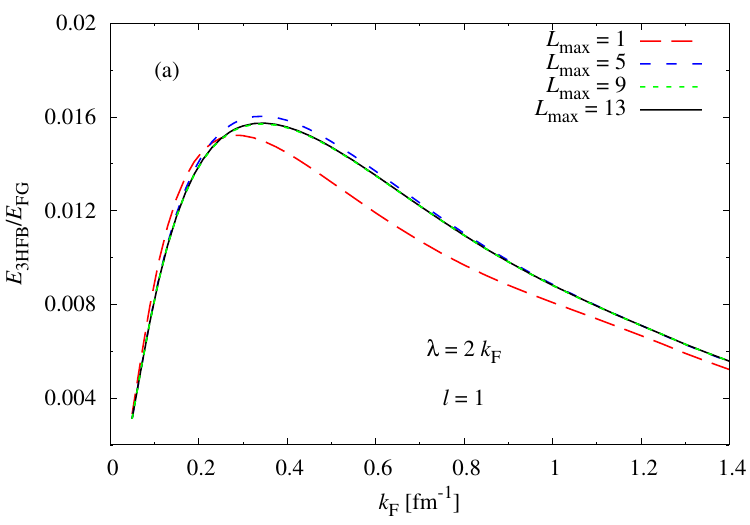}}
    {\includegraphics[width=\columnwidth]{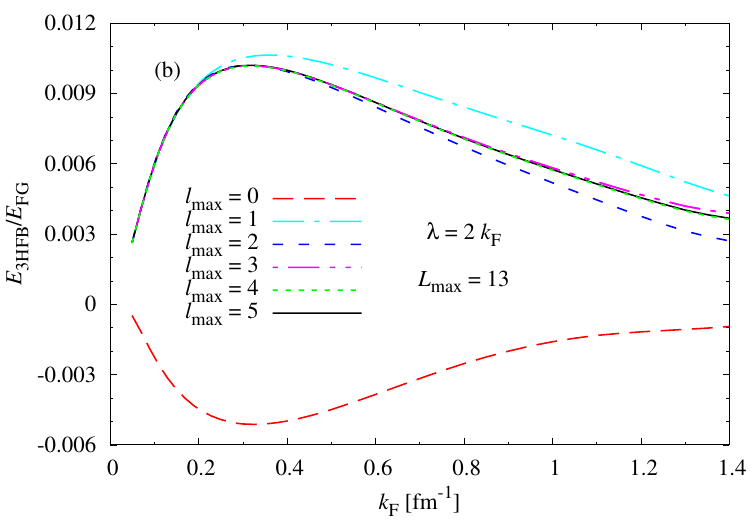}}
    \caption{Convergence of the HPW expansion of the HFB energy of the induced 3N interactions in units of the energy of the ideal Fermi gas as a function of the Fermi momentum.}
    \label{fig:e3hfb_conv}
\end{figure}

Coming back to the question of convergence, Fig.~\ref{fig:e3hfb_conv}(a) shows the convergence of the 3N contribution with respect to $L$ with a fixed $l=1$ 
and Fig.~\ref{fig:e3hfb_conv}(b) shows the convergence of the same with respect to $l$. 
The energy has converged with the inclusion of hyperspherical partial waves up to $L_{\maxi} = 13$ and $l_{\maxi} = 5$. 
It is interesting to note that $l=1$ is the dominant contribution, and not $l=0$. This is because $l=1$ contains $L=L'=1$, whereas $l=0$ starts only from $L=L'=2$ due to antisymmetrization, which forbids $L$ or $L'=0$. Also, note that odd and even $l$ contributions have opposite signs.

\begin{figure}
    \centering
    {\includegraphics[width=\columnwidth]{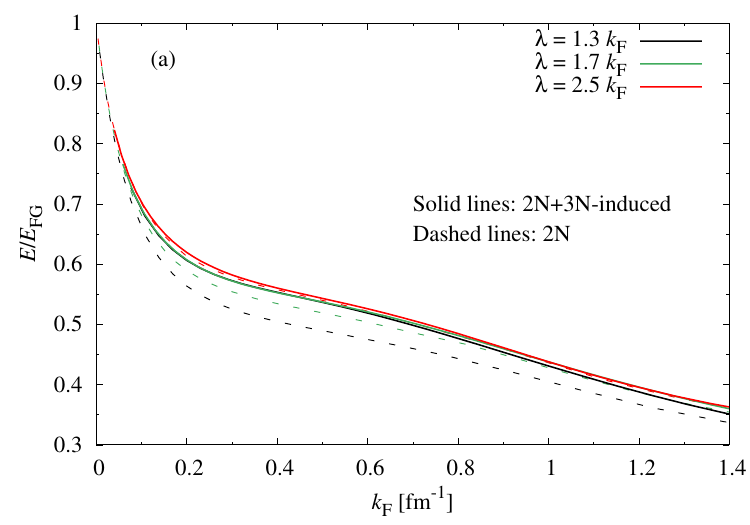}
    \includegraphics[width=\columnwidth]{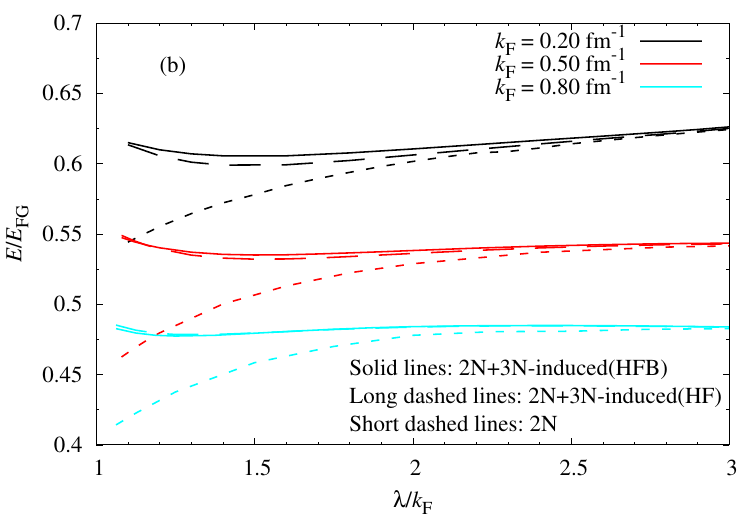}}
    \caption{Cutoff Dependence of the EoS of dilute neutron matter with the 2N (BMBPT3) interactions (short-dashed lines) only, with the 2N (BMBPT3) + induced 3N (HF) interactions (long-dashed lines) and with the 2N (BMBPT3) + induced 3N (HFB) interactions (solid lines): (a) The ground state energy as a function of the Fermi momenta for different values of the scale factor $f = \lambda/\kF$, (b) The ground state energy as a function of the scale factor $f = \lambda/\kF$ for different values of the Fermi momenta. 
    All energies are shown in units of the energy of the ideal Fermi gas.}
    \label{fig:eos_srg}
\end{figure}
\subsection{Cutoff dependence of the neutron-matter EoS}
We now look at the EoS of neutron matter, including the induced 3N contribution described in the preceding subsection, and in particular at the cutoff dependence of the results. 
Fig.~\ref{fig:eos_srg} shows the EoS of dilute neutron matter with the 2N interactions only and with the 2N + induced 3N interactions.
As mentioned earlier, the 2N calculations presented here are BMBPT3 (HFB + BMBPT up to third order) results obtained with the full SRG interactions.

In Fig.~\ref{fig:eos_srg}(a), we show the EoS as a function of the Fermi momentum for different scale factors $f=\lambda/\kF$.
Looking at the 2N results (dashed lines), we note that they present some cutoff dependence (i.e., dependence on the scale factor $f$).
The cutoff dependence gets strongly reduced when we add the HFB contribution of the SRG induced 3N interactions to the SRG 2N results (solid lines).
This confirms that the cutoff dependence of the 2N calculation stems mostly from the missing induced 3N interaction and not from missing higher orders of BMBPT, which is plausible because the third-order BMBPT correction is already very small in the chosen range of scale factors \cite{Palaniappan2023}.

In order to understand the cutoff dependence better, we plot in Fig.~\ref{fig:eos_srg}(b) the EoS as a function of the scale factor $f=\lambda/\kF$ for different Fermi momenta.
We see that within the 2N calculation, the energy strongly decreases with decreasing scale factor (short-dashed lines). 
Once we include the 3N contribution $\varepsilon_{3\text{HFB}}$ (solid line), this cutoff dependence is significantly reduced. 
Looking in detail, one sees that the correction computed in HFB, $\varepsilon_{3\text{HFB}}$, 
reduces the cutoff dependence slightly better than the one computed in HF, $\varepsilon_{3\text{HF}}$, (long-dashed lines).
The residual cutoff dependence of the EoS gets weaker as the Fermi momentum increases from $\kF =  0.2\, \fmi$  to $\kF =  0.8\, \fmi$.
However, notice that for $\kF = 0.8\,\fmi$ and $\lambda \gtrsim 2 \, \fmi$,
the matrix elements approach the initial ones, and $\lambda$ no longer reflects the true cutoff 
which can never become larger than the cutoff $2.5\,\fmi$ of the initial interaction.
For $\lambda/\kF\lesssim 1.3$, one may have to treat the induced 3N interaction in a more complete way than first-order perturbation theory, e.g., by including it into the self-consistent HFB calculation and into the BMBPT. 
In principle, it might be also necessary to include induced four-body interactions. 

Comparing Fig.~\ref{fig:eos_srg}(a) with Fig.~6 of our previous work \cite{Palaniappan2023} where we used $\Vlowk$ instead of SRG, we find that at the 2N level, the cutoff dependence of the EoS computed using the SRG-evolved interactions is more than twice as strong as the one of the EoS computed with the \Vlowk{} interactions.
This indicates that the 3N interactions induced by the \Vlowk{} flow are smaller than those generated by the SRG flow with the standard generator $\eta_s=[T,H_s]$.
But once the induced 3N contribution is added, the cutoff dependence of the 2N+3N SRG results is clearly weaker than the one of the 2N \Vlowk{} results.

\begin{figure}
    \centering
    {\includegraphics[width=\columnwidth]{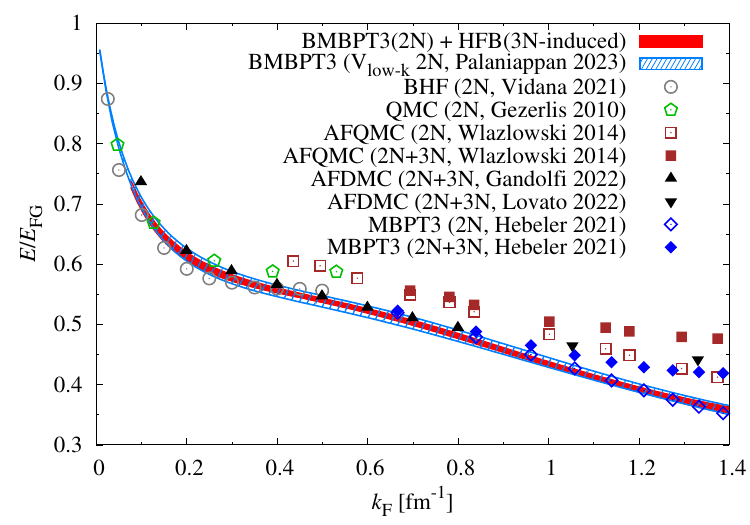}}
    \caption{BMBPT3(2N) + HFB(3N-induced) EoS of neutron matter within SRG (red band) compared with our previous BMBPT3(2N) \Vlowk{} results \cite{Palaniappan2023} (blue hatched band) and with other results from the literature: BHF (2N, Vida\~na 2021) \cite{Vidana2021}, QMC (2N, Gezerlis 2010) \cite{Gezerlis2010}, AFQMC (2N and 2N+3N, Wlaz\l{}owski 2014) \cite{Wlazlowski2014}, AFDMC (2N+3N, Gandolfi 2022) \cite{Gandolfi2022}, AFDMC (2N+3N, Lovato 2022) \cite{Lovato2022}, and MBPT3 (2N and 2N+3N, Hebeler 2021) \cite{Hebeler2021}.
    \label{fig:eos_pnm}}
\end{figure} 
This can be also seen in Fig.~\ref{fig:eos_pnm}. Since the residual cutoff dependence can be interpreted as a measure for the accuracy of the many-body calculation, we plot the results as bands that are obtained by varying the scale factor in a reasonable range ($1.3\leq \lambda/\kF\leq 2.5$). 
One can see that, except at very small values of $\kF$, the red band corresponding to the new 2N+3N SRG results is clearly narrower than the blue hatched band corresponding to the previous 2N \Vlowk{} results. 
Apart from that, the 2N + induced 3N SRG and the 2N \Vlowk{} results are in excellent agreement, although they are based on different initial 2N interactions (chiral N4LO in the case of SRG, AV18 in the case of \Vlowk). 
However, one should keep in mind that the error band obtained from the cutoff dependence does not take into account the uncertainties inherent to the initial interaction and due to missing physics, in particular, the contribution of the bare 3N interaction.

In Fig.~\ref{fig:eos_pnm}, we compare our results also with other results from the literature.
At low densities, we find our results to be in agreement with the Brueckner-Hartree-Fock (BHF) AV18 results from \cite{Vidana2021}, Quantum Monte Carlo (QMC) AV4 results from \cite{Gezerlis2010} (except for the last two points) and Auxiliary Field Diffusion Monte Carlo (AFDMC)  AV8'+UIX results from \cite{Gandolfi2022}.
At higher densities, we find agreement with 2N MBPT3 
(SRG $\lambda=2\,\fmi$) results from \cite{Hebeler2021} (Fig. 51).
However, our results lie below the 2N+3N calculations done using MBPT3 (SRG $\lambda=2\,\fmi$) \cite{Hebeler2021} (Fig. 51) and AFDMC AV18+UIX \cite{Lovato2022}. This hints at the importance of the bare 3N interaction at higher densities.
We note that the Auxiliary Field Quantum Monte Carlo (AFQMC) results of Ref.~\cite{Wlazlowski2014} (with and without 3N interaction) seem to be shifted
to higher energies compared to our and other results.

\section{Summary and Outlook}\label{sec:summary}
The present work is an extension of our preceding work \cite{Palaniappan2023} where we studied dilute neutron matter in the framework of BMBPT with $\Vlowk$ interactions 
evolved to very low cutoffs proportional to $\kF$. This method immediately gives rise to the question whether this approach can be trusted if the induced 3N interactions are 
neglected. Hence, it was necessary to compute the induced 3N interactions, for which the SRG offers a natural
framework. Our main result is that induced 3N interactions are 
indeed relevant for cutoffs less than $\sim 2\kF$ and that their inclusion reduces significantly the cutoff dependence. 
Furthermore, the final results for the equation of state (using SRG based on chiral N4LO interactions) agree very well with those obtained in our preceding work using $\Vlowk$ based on AV18, which shows the universality of the low-momentum effective interactions.

To achieve this, we computed the induced 3N interaction in the momentum space HPW basis using the SRG formalism. Since $s$-wave ($^1S_0$) is the dominant channel in dilute neutron matter, we limit ourselves to the 3N interaction induced by that channel.
We note that the convergence of the HPW expansion of the induced 3N matrix elements at low momenta in the SRG flow is improved if we avoid the intermediate summation in the $VV$ term by exploiting the spectator delta functions.

As long as the cutoff is not too small, the induced 3N contribution to the total energy is small and 
we can compute it in first-order perturbation theory by taking the expectation value in the HF or HFB ground state. 
The HF and HFB expectation values are repulsive and they are very similar to each other except at low densities.

When the SRG-evolved 2N interaction with density-dependent cutoff (i.e., cutoff scaled with the Fermi momentum)
is used as input, 
the EoS of dilute neutron matter within BMBPT shows a strong cutoff dependence when the cutoff gets close to $\kF$.
Once the induced 3N contribution is included, 
we find that this cutoff dependence reduces significantly.

Since the induced 3N interaction turns out to be important at low densities, 
it would be better to track its effect beyond first-order perturbation theory.
Its contribution to the EoS has to be properly studied by including it self-consistently into the HFB and also into the BMBPT calculation.
At higher densities, in addition to the 3N interactions induced by higher 2N partial waves, one needs to also include the bare 3N interaction \cite{Hebeler2021,Lovato2022}. We plan to address this problem in a future study.

Further, in addition to the EoS, the pairing gap is an important quantity for astrophysical purposes since it affects the thermal evolution and glitches of neutron stars. 
Therefore, it is important to address the problem of the pairing gap beyond the HFB level. We are currently working on a diagrammatic method to compute BMBPT corrections to the gap 
as the anomalous self-energy in the Nambu-Gor'kov formalism \cite{Schrieffer1964}. Screening and other corrections are then automatically included at higher orders. As 
for the EoS, it will be necessary to include not only the induced but also the bare 3N interactions into these calculations.

It would be also interesting to revisit the problem of the effective mass and the Landau parameters, since they are necessary to compute the linear response which is important e.g. in neutrino scattering but also for a more sophisticated treatment of the pairing with medium polarization \cite{Ramanan2018}. Compared to the effective mass that we compute in the 2N HFB \cite{Palaniappan2023}, there should be also corrections from the 3N interaction as well as higher orders in BMBPT. Concerning the Landau parameter $G_0$, it would be necessary to consider also spin-polarized matter.

However, in neutron stars, at densities above half the saturation density of symmetric nuclear matter, we are in the core where there is uniform matter with a finite proton fraction of a few percent. In that case, the dominant 3N interaction is the one between two neutrons and one proton since it is not suppressed by the Pauli principle.

Finally, let us mention that the approach of low-momentum interaction was also applied to systems of superfluid ultracold fermionic atoms \cite{Urban2021}. There, the induced 3N interaction was neglected. It would be interesting to include it and see if it helps to reduce the residual cutoff dependence as it does in the case of neutron matter. Work in this direction is in progress.

\begin{acknowledgments} 
We thank K. Hebeler for providing us with data from \cite{Hebeler2021} (Fig.~51).
We acknowledge support from the Collaborative Research Program of IFCPAR/CEFIPRA,
Project number: 6304-4.
\end{acknowledgments}
\appendix
\section{Normalization Convention for the Two-Body States}\label{appendix:2N_basis}
The 2N states in the center-of-momentum frame are  normalized as follows
\begin{gather}
    \braket{\vect{k}|\vect{k}'} = (2\pi)^3 \delta^{(3)}(\vect{k}-{\vect{k}'})\,,
    \nonumber\\
    \braket{\vect{k}|k'lm}
    = 2\pi^2 i^{-l} \frac{\delta(k-k')}{kk'} Y_{lm}(\hatvect{k})\,,
    \nonumber\\
    \braket{klm|k'l'm'}
    = \frac{\pi}{2} \frac{\delta(k-k')}{kk'} \delta_{ll'}\delta_{mm'}\,,
\end{gather}
 where $\vect{k}$ is the relative momentum,
$l$ and $m$ are the angular quantum numbers, and
$Y_{lm}$ is the spherical harmonic.

\section{Jacobi basis}\label{appendix:jacobi_basis}

Starting from the single-particle momenta $\{\vect{p}_1, \vect{p}_2, \vect{p}_3\}$, we can introduce the Jacobi system $\mathcal{J}^{(3)}\equiv\{\vect{P},\vect{k}^{(3)},\vect{q}^{(3)}\}$ via the following definitions of the so-called Jacobi momenta:
\begin{align}\label{Eq:JacobiDef}
    \vect{P} &= \vect{p}_1 + \vect{p}_2 + \vect{p}_3,
    \nonumber \\
    \vect{k}^{(3)} &= \frac{1}{2} (\vect{p}_2 - \vect{p}_1),
    \nonumber\\
    \vect{q}^{(3)} &= \frac{1}{\sqrt{3}}\bigg(\vect{p}_3-\frac{\vect{p}_1+\vect{p}_2}{2}\bigg).
\end{align}

The volume element transforms as the following
\begin{equation}
    d^3\vect{p}_1d^3\vect{p}_2d^3\vect{p}_3 = J \,d^3\vect{P}d^3\vect{k}^{(3)}d^3\vect{q}^{(3)},
\end{equation}
where $J = (2/\sqrt{3})^3$ is the Jacobian of the transformation. We can define Jacobi systems $\mathcal{J}^{(1)}$ and $\mathcal{J}^{(2)}$ similarly via cyclic permutations of $(123)$ in \Eq{Eq:JacobiDef}. 
In addition to these three systems, we can also define three more systems: $\mathcal{J}^{(\bar{i})}\equiv\{\vect{P},\vect{k}^{(\bar{i})},\vect{q}^{(\bar{i})}\}$ for $i=1,2,3$, where
\begin{equation}
    \vect{k}^{(\bar{i})}=-\vect{k}^{(i)},
    \hspace{1cm}
    \vect{q}^{(\bar{i})}=\vect{q}^{(i)}.
\end{equation}
Defining six such choices will make it easier to deal with the permutation of particles. 
These six different Jacobi systems are related through orthogonal transformations. For instance, the Jacobi systems $\mathcal{J}^{(j)}$ and $\mathcal{J}^{(\bar{i})}$ are related by
    \begin{equation}\label{eq:jacobi_diff}
        \begin{bmatrix}
            \vect{k}^{(\bar{i})} \\ \vect{q}^{(\bar{i})}
        \end{bmatrix}
        = 
        \begin{bmatrix}
            \cos{\theta_{ij}} & \sin{\theta_{ij}} \\
            \sin{\theta_{ij}} & -\cos{\theta_{ij}}
        \end{bmatrix}
        \begin{bmatrix}
            \vect{k}^{(j)} \\ \vect{q}^{(j)}
        \end{bmatrix},
    \end{equation}
where $\theta_{12}=\theta_{23}=\theta_{31}=\pi/3$, $\theta_{21}=\theta_{32}=\theta_{13}=-\pi/3$, and $\theta_{11}=\theta_{22}=\theta_{33}=\pi$.
The 3N states corresponding to the different Jacobi systems are defined as follows
\begin{widetext}
\begin{align}
    \ket{\vect{P},\vect{k},\vect{q}}^{(1)} 
    &\equiv
    \bigg|\frac{\vect{P}}{3}+\frac{2\vect{q}}{\sqrt{3}}\bigg\rangle_1\otimes
    \bigg|\frac{\vect{P}}{3}-\vect{k}-\frac{\vect{q}}{\sqrt{3}}\bigg\rangle_2\otimes
    \bigg|\frac{\vect{P}}{3}+\vect{k}-\frac{\vect{q}}{\sqrt{3}}\bigg\rangle_3,
    \nonumber\\
    \ket{\vect{P},\vect{k},\vect{q}}^{(2)} 
    &\equiv
    \bigg|\frac{\vect{P}}{3}+\vect{k}-\frac{\vect{q}}{\sqrt{3}}\bigg\rangle_1\otimes
    \bigg|\frac{\vect{P}}{3}+\frac{2\vect{q}}{\sqrt{3}}\bigg\rangle_2\otimes
    \bigg|\frac{\vect{P}}{3}-\vect{k}-\frac{\vect{q}}{\sqrt{3}}\bigg\rangle_3,
    \nonumber\\
    \ket{\vect{P},\vect{k},\vect{q}}^{(3)} 
    &\equiv
    \bigg|\frac{\vect{P}}{3}-\vect{k}-\frac{\vect{q}}{\sqrt{3}}\bigg\rangle_1\otimes
    \bigg|\frac{\vect{P}}{3}+\vect{k}-\frac{\vect{q}}{\sqrt{3}}\bigg\rangle_2\otimes
    \bigg|\frac{\vect{P}}{3}+\frac{2\vect{q}}{\sqrt{3}}\bigg\rangle_3,
    \nonumber\\
    \ket{\vect{P},\vect{k},\vect{q}}^{(\bar{i})} 
    &\equiv \ket{\vect{P},-\vect{k},\vect{q}}^{(i)} \hspace{2cm} (i=1,2,3).
\end{align}
\end{widetext}

The different states defined above are related by the permutations.
\begin{align}
    P^{(ij)}\ket{\vect{P},\vect{k},\vect{q}}^{(k)} &= \ket{\vect{P},\vect{k},\vect{q}}^{(\bar{k})},
    \nonumber\\
    P^{(ik)}\ket{\vect{P},\vect{k},\vect{q}}^{(k)} &= \ket{\vect{P},\vect{k},\vect{q}}^{(\bar{i})},
    \nonumber\\
    P^{(ijk)}\ket{\vect{P},\vect{k},\vect{q}}^{(k)} &= \ket{\vect{P},\vect{k},\vect{q}}^{(i)}
\end{align}
where $i\neq j\neq k$ and $P^{(ijk)}=P^{(ij)}P^{(jk)}$. Note that $P^{(ijk)} = P^{(jki)}= P^{(kij)}$.

From now on, for notational simplicity, 
we will suppress the superscript labeling of the Jacobi system 
whenever there is no ambiguity.

The normalization convention for the 3N states in a given Jacobi system is chosen as the following,
\begin{align}
    \braket{\vect{P}\vect{k}\vect{q}|\vect{P}'\vect{k}'\vect{q}'} = \frac{(2\pi)^9}{J} \delta(\vect{P}-{\vect{P}'}) \delta(\vect{k}-{\vect{k}'}) \delta(\vect{q}-{\vect{q}'}),
\end{align}
and the 3N states in the center-of-momentum frame are normalized as
\begin{align}
    \braket{\vect{k}\vect{q}|\vect{k}'\vect{q}'} = {(2\pi)^6} \delta(\vect{k}-{\vect{k}'}) \delta(\vect{q}-{\vect{q}'}).
\end{align}
Similar to the 2N partial wave basis,
we can expand $\ket{\vect{k}\vect{q}}$
using the spherical harmonics to get a 3N partial wave basis 
called the \textit{Jacobi partial wave basis} \cite{Hebeler2021}.

\section{Hyperspherical basis}\label{appendix:HPW_basis}
Combining the two Jacobi momenta in a six-dimensional hypermomentum vector $\vech{K} = (\kv, \qv)$, we can define a radial hypermomentum $K$ and a hyperangle $\alpha$ as follows
\begin{align}
    K = \sqrt{k^{2}+q^{2}},
    \hspace{1cm} 
    \alpha = \arccos\bigg( \frac{k}{K} \bigg),
\end{align}
with $K\geq 0$, and $\alpha\in[0,\frac{\pi}{2}]$.
Note that the radial hypermomentum $K$ is independent of the choice of the Jacobi system.
The hypermomentum vector $\vech{K}$ can then be
specified by its radial component $K$ 
and a set of five angular components 
$\{\alpha,\theta_{\vect{k}},\phi_{\vect{k}},\theta_{\vect{q}},\phi_{\vect{q}}\}$.
The volume element is then given by
\begin{align}
d^3\vect{k}d^3\vect{q}
= d^6 \vech{K}
= dK K^5 d^5 \hatvech{K},
\end{align}
where $d^5\hatvech{K} = \sin^2{\alpha} \,\cos^2{\alpha} \,d\alpha\,
d^2\hatvect{k}\,d^2\hatvect{q}$.

The ordinary three-dimensional spherical harmonics can be generalized to $d$-dimensional hyperspherical harmonics (HH).
The six-dimensional HH is given by \cite{Avery2018,Barnea2022}
\begin{align}\label{eq:RRC}
    \mathcal{Y}_{[L]}(\hatvech{K}) 
    = Y_{l_1 m_1}(\hatvect{k})\,Y_{l_2 m_2}(\hatvect{q}) \, \mathcal{P}_L^{l_2,l_1}(\alpha),
\end{align}
Here, $[L]\equiv\{L,l_1,m_1,l_2,m_2\}$ denotes a set of five quantum numbers.
$L$ is called the \textit{grand orbital quantum number}.
$L\geq l_1+l_2$ and must have the same parity as $l_1+l_2$.
$\mathcal{P}_L^{l_2,l_1}(\alpha)$ is given by
\begin{align}
    \mathcal{P}_L^{l_2,l_1}(\alpha)
    = N_L^{l_2,l_1} (\sin{\alpha})^{l_2} (\cos{\alpha})^{l_1} P^{(l_2+\frac{1}{2},l_1+\frac{1}{2})}_{n}(\cos{2\alpha}),
\end{align}
where $P^{(l_2+\frac{1}{2},l_1+\frac{1}{2})}_{n}$ is the Jacobi polynomial, $n={(L-l_1-l_2)}/{2}$ is a non-negative integer
and $N_L^{l_2l_1}$ is a normalization constant given by 
\begin{equation}
N_L^{l_2l_1}
=\sqrt{
\frac{2(L+2)\Gamma(n+1)\Gamma(n+l_1+l_2+2)}
{\Gamma(n+l_1+\frac{3}{2})\Gamma(n+l_2+\frac{3}{2})}}.    
\end{equation}

We can then expand the ket $\ket{\vech{K}}$ using the HH to get the \textit{hyperspherical partial wave basis}.
The HPW states are normalized as follows,
\begin{align}\label{Eq:HPW_basis}
    \braket{\vech{K}|K'[L]}
    &= (2\pi)^3 i^{-L} \frac{\delta(K-K')}{\sqrt{(KK')^5}} \mathcal{Y}_{[L]}(\hatvech{K}),
    \nonumber\\
    \braket{K[L]|K'[L']}
    &= \frac{\delta(K-K')}{\sqrt{(KK')^5}}\delta_{[L][L']}.
\end{align}

The HH corresponding to the coupled quantum numbers is given by
\begin{align}
    \mathcal{Y}_{[L]_c}(\hatvech{K}) 
    &= \sum_{m_1,m_2} C^{lm}_{l_1m_1l_2m_2}\mathcal{Y}_{[L]}(\hatvech{K}),
\end{align}
where ${[L]}_c\equiv\{L,l,m,l_1,l_2\}$, 
$C^{lm}_{l_1m_1l_2m_2}$ is the Clebsch-Gordon coefficient. 
The normalization of the HPW states in the coupled basis can be obtained by simply replacing $[L]$ with $[L]_c$ in \Eq{Eq:HPW_basis}.

\begin{figure}
\centering
\includegraphics{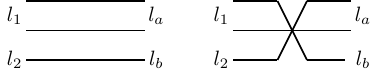}
\caption{Schematic representation of the two terms in the coefficient $\mathcal{A}^{Ll}_{l_1l_2l_al_b}(\theta_{13})$.}
\label{fig:schematic-A}
\end{figure}

The HH corresponding to different Jacobi systems are related through the \textit{Reynal-Revai coefficients} (RR coefficients) \cite{Raynal1970, Efros2020, Efros2021} For instance, given two Jacobi systems related through \Eq{eq:jacobi_diff}, the corresponding HH are related by
    \begin{align}\label{eq:rrcoeff1}
        \mathcal{Y}_{{[L_{(a)}]}_c}(\hatvech{K}^{(\bar{i})}) 
        = \sum_{l_1,l_2} \; \langle l_1\, l_2 | l_a \, l_b \rangle^{\theta_{ij}}_{Ll} \;
        \mathcal{Y}_{{[L]}_c}(\hatvech{K}^{(j)}),
    \end{align}
where ${[L_{(a)}]}_c\equiv\{L,l,m,l_a,l_b\}$ and the term with the angle bracket is the RR coefficient\footnote{We use an efficient code by Efros for computing the RR coefficients \cite{Efros2021}. Since our convention for the hyperangle $\alpha$ is different from Efros' one (who uses $\frac{\pi}{2}-\alpha$ instead), we have to multiply the RR coefficients obtained from his code by $(-1)^{(l_1+l_2-l_a-l_b)/2}$. This is because $\mathcal{P}^{l_2l_1}_L(\frac{\pi}{2}-\alpha)=(-1)^{n}\mathcal{P}^{l_1l_2}_L(\alpha)$, with $n = (L-l_1-l_2)/2$.}.
The action of the operator $A^{(ij)}$ on the hypermomentum state is given by
\begin{align}
    A^{(ij)} \ket{\vech{K}}^{(j)} 
    &= (2\pi)^3 i^{L} \sum_{[L]_c} \mathcal{Y}^{*}_{[L]_c}(\hatvech{K}^{(j)}) 
        \nonumber\\
        &\hspace{0.5cm}\times\bigg[ \sum_{l_al_b} \mathcal{A}^{Ll}_{l_1l_2l_al_b}(\theta_{ij}) \ket{K[L_{(a)}]_c}
        \ \bigg]
\end{align}
where $\mathcal{A}^{Ll}_{l_1l_2l_al_b}(\theta_{ij})=\delta_{l_1l_a}\delta_{l_2l_b}-\braket{l_1l_2|l_al_b}^{\theta_{ij}}_{Ll}$ 
(see Fig.~\ref{fig:schematic-A} for a schematic representation).

\section{$VV$ term without the intermediate $L''$ summation}\label{appendix:VV_term}
\begin{figure}
    \centering
    \includegraphics{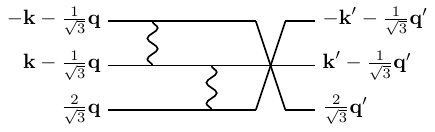}
    \caption{Schematic representation of \Eq{eq:VVP} in the third Jacobi system.}
    \label{fig:VVP}
\end{figure}

In order to get an expression for the VV term without the intermediate $L''$ summation, 
write in Eq.~\eqref{eq:srg3sp} $V^{(32)}A^{(13)} = - V^{(32)}P^{(13)}A^{(13)}$. 
Hence, up to a minus sign, the product of potentials in the $VV$ term before antisymmetrization becomes 
\begin{align}
    &V^{(12)}_{123789}(V^{(32)}P^{(13)})_{789456}
    \nonumber\\
    &\quad= V^{(12)}_{123789}V^{(32)}_{789654}
    \nonumber\\
    &\quad= V_{1278}\delta_{39}V_{9845}\delta_{76}
    \nonumber\\
    &\quad=\Big(\frac{4\pi}{\calV}\Big)^2\, V\Big(\frac{|\vect{p}_2-\vect{p}_1|}{2},\frac{|\vect{p}_8-\vect{p}_7|}{2}\Big) \delta_{1+2,7+8}\delta_{39}
    \nonumber\\
    &\quad\quad \times V\Big(\frac{|\vect{p}_8-\vect{p}_9|}{2},\frac{|\vect{p}_5-\vect{p}_4|}{2}\Big)\delta_{8+9,4+5}\delta_{76}.
    \label{eq:VVP}
\end{align}
Using the delta functions, one gets $\vect{p}_7 = \vect{p}_6$, $\vect{p}_8 = \vect{p}_1+\vect{p}_2-\vect{p}_6$, $\vect{p}_9 = \vect{p}_3$ along with total momentum conservation. Expressing $123$ and $456$ in the third Jacobi system as $\vect{k}$, $\vect{q}$ and $\vect{k}'$, $\vect{q}'$, respectively (see Fig.~\ref{fig:VVP}), one gets
\begin{align}
   &V\Big(\frac{|\vect{p}_2-\vect{p}_1|}{2},\frac{|\vect{p}_8-\vect{p}_7|}{2}\Big) 
   V\Big(\frac{|\vect{p}_8-\vect{p}_9|}{2},\frac{|\vect{p}_5-\vect{p}_4|}{2}\Big) 
   \nonumber\\
   &\quad\quad= 
   V\Big(k,\frac{|\vect{q}+2\vect{q}'|}{\sqrt{3}}\Big)
   V\Big(\frac{|\vect{q}'+2\vect{q}|}{\sqrt{3}},k'\Big).
\end{align}

Similarly, one finds for the energy difference that appears in the $VV$ term of Eq.~\eqref{eq:srg3sp}:
\begin{equation}
    \varepsilon_{123}+\varepsilon_{456}-2\varepsilon_{789} = k^2+q^2+k'^2+q'^2-\frac{8}{3}(q^2+q'^2+\vect{q}\cdot\vect{q}').
\end{equation}

Finally, one has to project on one $lLL'$ channel. Hence, including all necessary factors, one gets \Eq{eq:VV_HPW}.
\bibliography{references}
\end{document}